\documentclass[fleqn,10pt]{wlscirep}
\usepackage[utf8]{inputenc}
\usepackage[T1]{fontenc}
\usepackage{graphicx}
\usepackage{dcolumn}
\usepackage{lscape}
\usepackage{booktabs}
\usepackage{longtable}
\usepackage{etoolbox}
\BeforeBeginEnvironment{longtable}{\begin{center}\small}
\AfterEndEnvironment{longtable}{\end{center}}
\usepackage{bm}

\usepackage{hyperref}
\hypersetup{
    colorlinks=true,
    linkcolor=blue,
    citecolor=blue,
    filecolor=cyan,
    urlcolor=blue,
}
\usepackage[utf8]{inputenc}
\usepackage[T1]{fontenc}
\usepackage{mathptmx}
\usepackage{wasysym}
\title{PaleoJump: A database for abrupt transitions in past climates}

\author[1,*]{Witold Bagniewski}
\author[2,3,4]{Denis-Didier Rousseau}
\author[1,5]{Michael Ghil}
\affil[1]{Laboratoire de Météorologie Dynamique, École Normale Supérieure and PSL University, Paris, France}
\affil[2]{Geosciences Montpellier, University of Montpellier, CNRS, Montpellier, France}
\affil[3]{Institute of Physics - CSE, Division of Geochronology and Environmental Isotopes, Silesian University of Technology, Gliwice, Poland}
\affil[4]{Lamont-Doherty Earth Observatory, Columbia University, New York, USA}
\affil[5]{Department of Atmospheric and Oceanic Sciences, University of California at Los Angeles, Los Angeles, USA}
\affil[*]{wbagniewski@lmd.ipsl.fr}

\begin{abstract}
Tipping points (TPs) in the Earth system have been studied with growing interest and concern in recent years due to the potential risk of anthropogenic forcing causing abrupt, and possibly irreversible, climate transitions. Paleoclimate records are essential for identifying TPs in Earth’s past and for properly understanding the climate system’s underlying nonlinearities and bifurcation mechanisms. Due to the variations in quality, resolution, and dating methods, it is crucial to select the records that give the best representation of past climates. Furthermore, as paleoclimate time series vary in their origin, time spans, and periodicities, an objective, automated methodology is crucial for identifying and comparing TPs. To reach this goal, we present here the PaleoJump database of carefully selected, high-resolution records originating in ice cores, marine sediments, speleothems, terrestrial records, and lake sediments. These records describe climate variability on centennial, millennial, or longer time scales and cover all the continents and ocean basins. We provide an overview of their spatial distribution and discuss the gaps in coverage. Our statistical methodology includes an augmented Kolmogorov-Smirnov test and Recurrence Quantification Analysis; it is applied here to selected records to automatically detect abrupt transitions therein and to investigate the presence of potential tipping elements. These transitions are shown in the PaleoJump database together with other essential information, including location, temporal scale and resolution, along with temporal plots. This database represents, therefore, a valuable resource for researchers investigating TPs in past climates.
\end{abstract}
\begin{document}

\flushbottom
\maketitle

\thispagestyle{empty}


\section*{Introduction and Motivation} \label{sec:intro}

Ever since Dansgaard–Oeschger events were discovered in ice cores from Greenland \cite{dansgaard1982,johnsen1992,grootes1993}, climate research has aimed to identify other examples of centennial-to-millennial climate variability, including in marine and terrestrial paleoclimate records \cite{shackleton2000,genty2003}, and to gain insight into their mechanisms. Many such records have been found to exhibit abrupt transitions, raising the question of whether similar drastic changes may occur in the nearby future, as anthropogenic global warming is pushing the climate system away from the relatively stable state that has persisted throughout the Holocene. Many of Earth's subsystems exhibit intrinsic variability and respond nonlinearly to various forcings, both natural and anthropogenic \cite{Ghil.Chil.1987,Ghil.Lucar.2020}. Hence, any of these subsystems could abruptly shift into a new state once particular key thresholds, known as tipping points (TPs), are crossed \cite{pearce2007,lenton2008,scheffer2009}.

Identifying potential TPs in the climate system requires theoretical and modelling work, including comparison with observations. Proxy records of past climates play a crucial role, by enabling the reconstruction of Earth's climatic history. Numerous well dated high-resolution records include abrupt transitions and may thus give insights into TPs in the Earth system’s past. As the number of available paleoproxy datasets is in the tens of thousands, finding and selecting the records that are most relevant for studying TPs is a daunting task. These proxy datasets originate from different geologic structures, contain different variables, and span a wide range of age intervals with different resolutions.

Thus, a comprehensive database of paleoclimate records that contain abrupt transitions can provide valuable information for identifying critical TPs in current and future climate evolution. Furthermore, as tipping elements --- i.e., the subsystems that may be subject to tipping --- are interconnected, a potential for domino effects exists \cite{wunderling2021}. To identify and describe such effects from past records, one needs coverage from different types of archives with a comprehensive geospatial distribution.

The purpose of this paper is to address these challenges by presenting the PaleoJump database \cite{BRG_PJ}, \href{https://paleojump.github.io}{https://paleojump.github.io}, which compiles globally sourced high-resolution paleoclimate records originating in ice, marine sediments, speleothems, terrestrial deposits, and lake sediments. The database is designed as a website, allowing easy access and navigation. It includes a map of the paleoclimate records, as well as tables that list supplementary information for each record.

Since paleoclimate records vary in their origin, time spans, and periodicities, an objective, automated methodology is key for identifying and comparing TPs. Here, we apply a recently developed method to detect abrupt transitions based on an augmented Kolmogorov-Smirnov (KS) test \cite{bagniewski2021} to selected records within the database. The KS results are compared with recurrence quantification analysis (RQA) \cite{eckmann1987,marwan2007}.

\section*{Database sources} \label{sec:data}

The PaleoJump database currently includes records from 123 sites, grouped by their geological type: 49 marine-sediment cores, 29 speleothems, 18 lake sediment cores, 16 terrestrial records, and 11 ice cores. The main sources for this database are the PANGAEA and NCEI/NOAA open-access data repositories, while some records are, unfortunately, available only as supplementary files of the articles describing the records; in the latter case, links to the corresponding articles are provided. The paleorecords have been selected for their ability to represent different aspects of past climate variability and are characterized by high temporal resolution, multi-millennial time scales, and a comprehensive spatial coverage. This selection simplifies the search for records that are most helpful in the investigation of critical transitions and of the behavior of tipping elements.

While many of the paleosites included in PaleoJump include multiple proxy types, we have focused on proxies that can be directly compared with climate models: oxygen isotopes reflecting changes in past temperatures, sea level, and precipitation; carbon isotopes containing information on past vegetation and the carbon cycle; aeolian deposits that include signatures of past precipitation, mineral aerosols, and atmospheric transport patterns; as well as other proxy-based estimates of past temperatures. We have mainly focused on the Last Climate Cycle, due to the well-established evidence of past abrupt transitions --- such as Dansgaard-Oeschger (DO) and Heinrich events --- with most records also covering Holocene deglaciation. Other records extend further back in time, including DO-like events during earlier glacial cycles of the Quaternary, and earlier climatic events of the Cenozoic era, such as the Eocene--Oligocene Transition at 34~Ma or the Paleocene-Eocene Thermal Maximum (PTEM) at 56~Ma. While PaleoJump provides global coverage with records from all continents and ocean basins, its spatial coverage is biased towards the North Atlantic region due to greater availability and a strong impact of the DO events.

Five tables show the information for each record in the PaleoJump database and are included in the SM \cite{bahr2018,barker2014,barker2019,beuscher2017,bolton2018,clemens2018,clemens2021,davtian2021,deabreu2003,dedeckker2020,deplazes2013,deplazes2014,depolholz2007,dickson2008,dokken1999,ehrmann2021,elderfield2012,eynaud2009,gottschalk2015,harada2006,hendy1999,hendy2002,hendy2003,hodell2003,hodell2008,hodell2010,hodell2013,hodell2015,hodell2016,jung2009,lauterbach2020,lea2006,martrat2007,mohtadi2014,naafs2013,naafs2020,naughton2009,nurnberg2008,pahnke2003,pichevin2007,rampen2012,rickaby1999,riveiros2010,rosenthal2003,saikku2009,salgueiro2014,sanchez2017,schulz1998,shackleton2000,stott2002,vankreveld2000,voelker2006,voelker2010,waelbroeck2019,weijers2009,weldeab2003,weldeab2007,westerhold2020,zarriess2010,zhao1995,ziegler2013,barbante2006,barker2011,baumgartner2014,bazin2013,extier2018,jouzel2007,lambert2008,lambert2012,loulergue2008,luthi2008,petit1999,rasmussen2013,rasmussen2014,thompson1997,vallelonga2013,uemura2018,wais2015,arienzo2015,asmerom2010,bar2003,boch2011,burns2003,carolin2013,cheng2012,cheng2016,cruz2005,denniston2013,fleitmann2009,genty2003,genty2010,kanner2012,kathayat2016,kelly2006,lachniet2009,lachniet2014,mosblech2012,moseley2016,moseley2020,unal2015,wagner2010,wang2001,wang2008,wang2017,whittaker2011,hao2012,moine2017,seelos2009,sun2010,sun2012,rousseau2017,ujvari2014,yang2014,allen1999,allen2000,ampel2008,benson2003,camuera2018,donders2021,follieri1989,fritz2007,fritz2010,grimm2006,huntley1999,johnson2016,melles2012,meyer2014,miebach2019,muller2003,pickarski2017,prokopenko2006,reille1990,sadori2016,tierney2008,tzedakis2004,tzedakis2006,veres2009,wagner2019}.

\begin{figure*}[!htbp]
\includegraphics[width=1\linewidth]{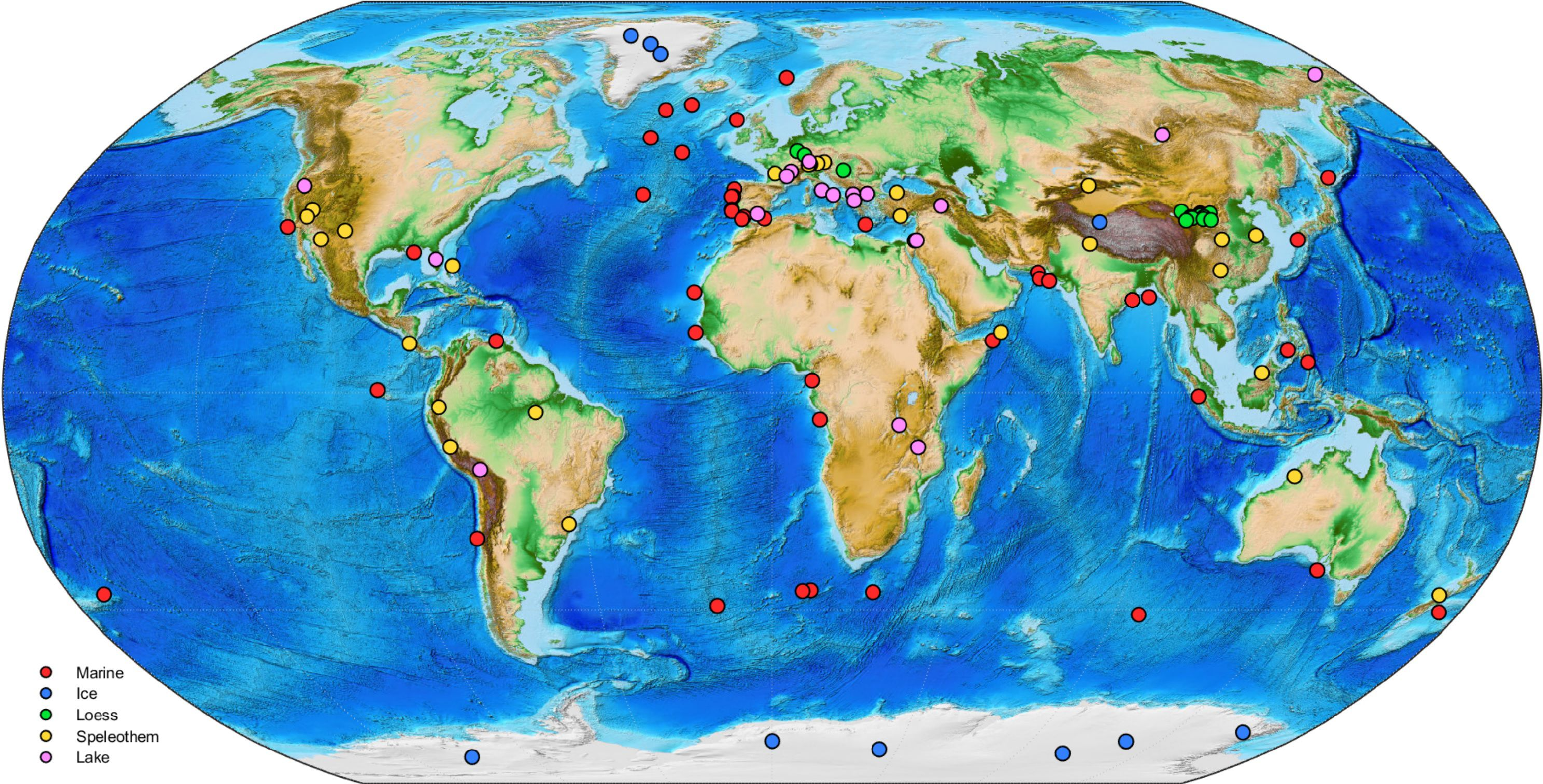}
    \caption{Map of records listed in the Supplementary Material (SM) and included in the PaleoJump database. The five record sources --- marine sediments, ice cores, terrestrial deposits, speleothems, and lake sediments --- are identified in the legend by color.}
\label{map}
\end{figure*}

\section*{Applying the KS test and RQA to TP identification} \label{sec:KS_test}

Given the diversity of paleoclimate records, an objective, automated methodology is crucial for identifying and comparing TPs. Bagniewski et al. \cite{bagniewski2021} have formulated an augmented KS test and applied it successfully to the robust detection and identification of abrupt transitions for the last glacial cycle. Their results were compared with RQA, showing the complementarity of the two methods, with KS more useful at detecting individual jumps and finding their precise dates, while RQA can help establish important transitions in a record’s characteristic time scale.
Here we apply these two methods to selected records of the PaleoJump database and demonstrate the ability of the KS test to accurately identify transitions for different types of paleoclimate records.

\paragraph{Kolmogorov-Smirnov (KS) test.}

The augmented KS methodology \cite{bagniewski2021} is based on the nonparametric KS test \cite{Massey.1951}. A two-sample KS test is applied to two neighboring samples drawn from a proxy time series within a sliding window of length $w$. The commonality of the two samples is quantified by the the KS statistic $D_{\rm{KS}}$ \cite{Massey.1951,conover1999}. A ``jump'' in the time series is identified at any point in time at which $D_{\rm{KS}}$ is greater than a cut-off threshold $D_{\rm c}$. As the KS test can give very different results depending on the window length $w$ being used, $D_{\rm{KS}}$ is calculated for different $w$'s, varying between $w_{\rm{min}}$ and $w_{\rm{max}}$. The values of the latter two parameters bracket the desired time scale at which a given paleorecord is to be investigated. Furthermore, smaller jumps in the time series may be the result of an error in the observed data or small-scale variability that occurs over time intervals shorter than the sampling resolution of the proxy record and they should be discarded. Thus, for a transition to be considered significant, the change in magnitude between the two samples $(i, j)$ should exceed a threshold $\sigma_{\rm c}$ in their standard deviations $(\sigma_i, \sigma_j)$. Finally, as the KS test requires a large enough sample size to be significant, its results are rejected if either of the two samples has a size $n$ smaller than $n_{\rm c}$.

At a time step at which all three conditions based on the parameters $D_{\rm c}$, $\sigma_{\rm c}$, and $n_{\rm c}$ are satisfied, an abrupt transition is identified. As the dates of such transitions often occur in clusters, the precise date for a transition within such a cluster is determined by the maximum $D_{\rm{KS}}$ value found within the corresponding time interval. When the maximum $D_{\rm{KS}}$ over a given interval is shared by several time steps, the one corresponding to the maximum change in absolute magnitude is used; moreover, if there are several jumps of equal amplitude, then the one with the earliest date is used.

As the same transition may be found at slightly different dates depending on the window length that is used, we first identify the transitions detected with the longest window, which, given the larger sample size it accomodates, is the most statistically significant one. These transitions are then supplemented by those detected for the next-longest window and eventually for all other window lengths. For window $w_i$, we discard transitions identified at time $t$ if the interval $\{t - w_i \le t \le t + w_i\}$ contains transitions that were previously identified with a greater window length. Finally, long-term trends in maxima and minima are used to establish the main transitions, such as Stadial-Interstadial (GS -- GI) boundaries.

The parameters $D_{\rm c}$ and $\sigma_{\rm c}$ are initially optimized following receiver operating characteristic analysis \cite{fawcett2006,hastie2009}, and abrupt transitions so identified for the NGRIP ice core $\delta^{18}$O record are further compared with the change points identified using visual inspection by Rasmussen \emph{et al.} \cite{rasmussen2014}.

\paragraph{Recurrence Quantification Analysis (RQA).} \label{ssec:RQA}

The KS test results are next compared with RQA results \cite{eckmann1987,marwan2007,marwan2013}. Here, the Recurrence Plot (RP) for a time series $\{x_k: k = 1, \ldots, K\}$ is given by a square pattern in which both axes represent time. A dot is entered into a position $(i,j)$ of the matrix $\mathbf R$ when $|x_i - x_j| < \varepsilon$, with $\varepsilon$ being the recurrence threshold. Thus, the RP appears as a square matrix $\mathbf R$ of dots. For details on how $\varepsilon$ is determined, see Marwan \emph{et al.} \cite{marwan2007} and Bagniewski \emph{et al.} \cite{bagniewski2021}.

Eckmann et al. \cite{eckmann1987} showed that purely visual RP typologies provide useful information about a time series. However, RQA allows for a more objective way of inferring recurrence \cite{marwan2007, marwan2013}, by quantifying selected recurrence characteristics. One of the simplest RQA criteria is the recurrence rate (RR), namely the density of dots within the recurrence plot: RR describes the probability of states of the system recurring within a particular time interval. By evaluating RQA measures such as RR in a sliding window, it is possible to identify changes in the time series. Low RR values correspond to an unstable behavior of the system, and hence abrupt transitions in a time series may be identified by local RR minima.

An important advantage of the recurrence method is that it does apply to dynamical systems that are not autonomous, i.e., that may be subject to time-dependent forcing \cite{eckmann1987}. The latter is certainly the case for the climate system in general \cite{GCS.2008, CSG.2011, Bodai.Tel.2012, Ghil.2019} and, in particular,
on the time scales of 10–100~kyr and longer, which are affected strongly by orbital forcing \cite{Riechers.ea.2022}.

For a more comprehensive description of both the augmented KS test and RQA, see Bagniewski et al. \cite{bagniewski2021}.

\section*{Examples of usage} \label{sec:apply}

Here we show the results of the augmented KS test methodology \cite{bagniewski2021}, as applied to records of different timescales, resolutions, and periodicities. Plots of the detected transitions, along with spreadsheets listing their dates are available on the PaleoJump database for other records as well.

\paragraph{Methodology.} \label{ssec:methods}

To demonstrate the applicability of the PaleoJump database to the study of climate TPs, we tested herein the ability of the augmented KS test to detect abrupt transitions in different types of paleoproxy records. Specifically, we analyze six records, given in Table~\ref{tab:tableshort}, from each of the proxy types listed in Supplementary Tables 1 -- 5. In addition to these six records, we include the results obtained for the NGRIP ice core, which have been published in 
Bagniewski et al. \cite{bagniewski2021}, and compare the latter with the three records of the last climate cycle in the table, to wit MD03-2621, Paraiso Cave, and ODP893A.

\begin{longtable}{@{}lllllll@{}}
\toprule
\textbf{Type} & \textbf{Site name}  & \textbf{Location} & \textbf{Depth/elevation} & \textbf{Age} & \textbf{Res.} & \textbf{Proxy}                                                                                                                                      \\* \midrule
\endhead
\bottomrule
\endfoot
\endlastfoot
Marine      & ODP893A \cite{hendy2002,hendy1999}    & 34.28, -120.03    & 576 m     & 65 - 0 ka     & 41 y      & pla $\delta^{18}$O \\
Marine      & MD03-2621 \cite{deplazes2013}         & 10.678, -64.972   & 847 m     & 109 - 6 ka    & 0.1 y     & reflectance \\
Marine      & U1308 \cite{hodell2016}               & 49.878, -24.238   & 3871 m    & 3143 - 0 ka   & 118 y     & ben $\delta^{18}$O \\
Marine      & CENOGRID \cite{westerhold2020}        & N/A               & N/A       & 67.1 - 0 Ma   & 2000 y    & ben $\delta^{18}$O \\
Terrestrial & Paraiso (PAR07) \cite{wang2017}       & -4.067, -55.45    & 60 m      & 45 - 18 ka    & 21 y      & $\delta^{18}$O \\
Terrestrial & Lake Ohrid  \cite{sadori2016}         & 41.049, 20.715    & 693 m     & 1.36 - 0 Ma   & 208 y     & TIC \\
Ice         & NGRIP \cite{rasmussen2014}            & 75.1, -42.32      & 2925 m    & 122 - 0 ka    & 20 y      & $\delta^{18}$O \\* \bottomrule
\caption{Records analyzed using the augmented KS test \cite{bagniewski2021}, ordered by geological type and temporal scale; the nature of the records is indicated by the abbreviations ben = benthic, pla = planktonic, and TIC = Total Inorganic Carbon in the last column.}
\label{tab:tableshort}\\
\end{longtable}

\vspace{-1.0cm}
The KS test parameters vary depending on a record's time resolution and on the length of its age interval. For the records covering the last climate cycle (MD03-2621, Paraiso Cave, and ODP893A), we use the same parameter values as used for the NGRIP record in Bagniewski et al. \cite{bagniewski2021}, i.e., $D_{\rm c} = 0.77$, $\sigma_{\rm c} = 1.9$, $n_{\rm c} = 3$, $w_{\rm{min}} = 0.1$~kyr, and $w_{\rm{max}} = 2.5$~kyr. For the records spanning longer time intervals with a lower temporal resolution, we use longer window lengths $w_{\rm{min}}$ and $w_{\rm{max}}$, thus shifting the focus of our analysis to longer time scales. For the U1308 benthic $\delta^{18}$O and Lake Ohrid TIC records, we use a $w$-range of 2~kyr to 20~kyr. This allows us to focus on the glacial-cycle variability, as the record's resolution is too low to properly identify DO events, particularly for data older than 1.5~Ma~BP, when the U1308 record's resolution is lower than for more recent data. For the CENOGRID record, we perform two separate analyses, one with a $w$-range of 1~Myr to 4~Myr to determine the Quaternary's major climatic shifts, and one with a $w$-range of 0.02~Myr to 2.5~Myr, which covers the orbital time scale.\footnote{\label{BP} Note that, in paleoclimate studies, one distinguishes between units of absolute time, such as kyr or Myr, and units of age, such as ka~BP or Ma~BP, where `BP' stands for ``before present.''}

\paragraph{Results for Individual records.} \label{ssec:results1}

\begin{figure*}[!htbp]
\includegraphics[width=0.8\linewidth]{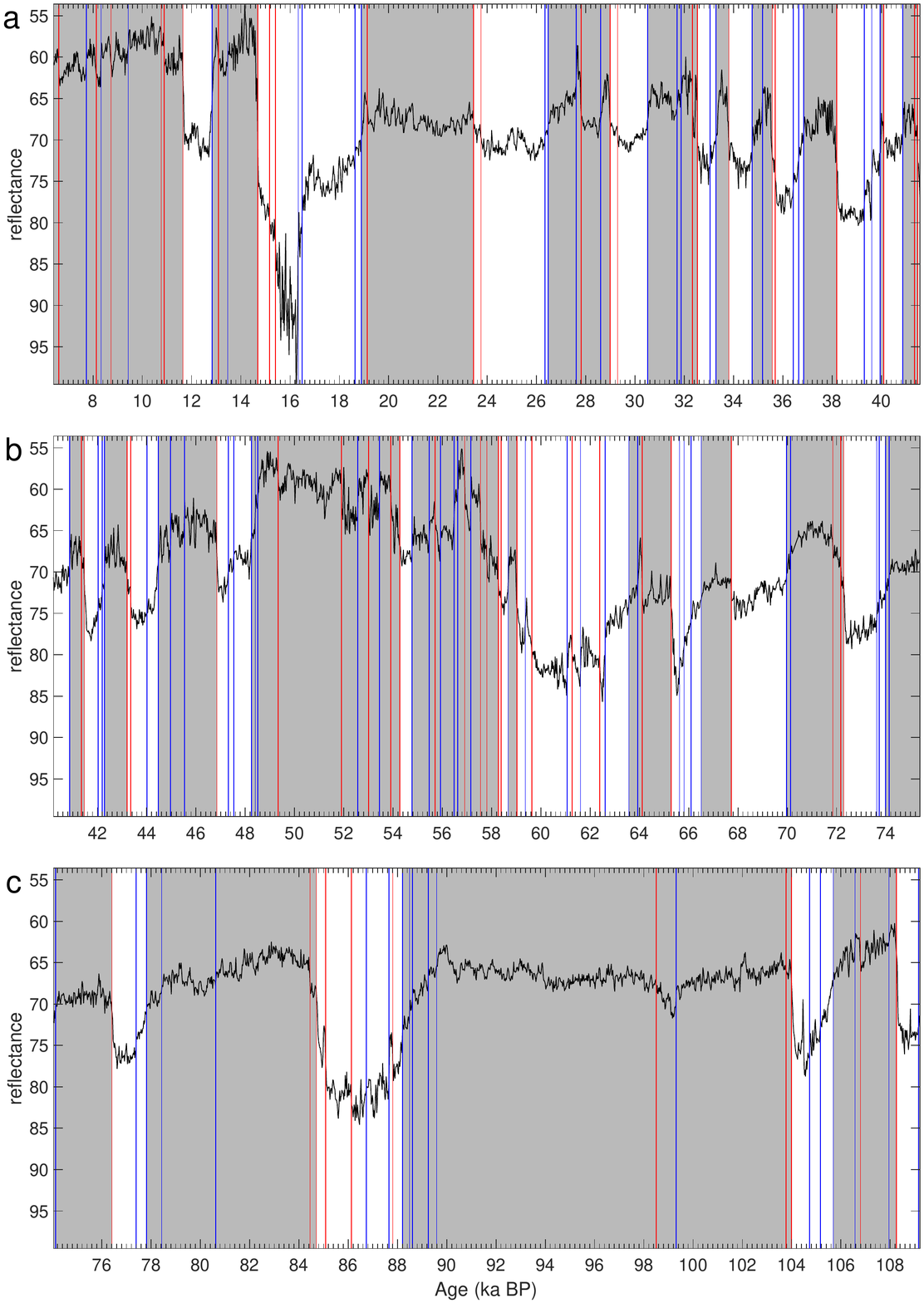}
    \caption{MD03-2621 marine sediment core. The black solid line is the core reflectance \cite{deplazes2013}, with the vertical axis reversed. Vertical lines represent transitions detected by the KS test of Bagniewski et al. \cite{bagniewski2021}, with red lines for warming transitions, i.e., DO events, and blue lines for cooling ones. Grey bars represent warm episodes. Note that all time axes in this paper follow the geological custom of pointing into the past.}
\label{MD03-2621}
\end{figure*}

We chose the MD03-2621 reflectance record from the Cariaco Basin in the Caribbean for having a very high resolution and for its importance in studying the effect of DO events on the migrations of the intertropical convergence zone (ITCZ). The record is shown in Fig.~\ref{MD03-2621} and it has been used previously to assess teleconnections between the North Atlantic basin and the Arabian Sea \cite{deplazes2013}. When the ITCZ migrates southward during stadials, northeasterly Trade winds lead to upwelling of cool, nutrient-rich waters in the Caribbean; as the ITCZ migrates northward during interstadials, heavy convective rainfall leads to increased runoff from South America's north coast, delivering detrital material to the Cariaco Basin \cite{deplazes2013,bradley2021}. As a result, the color reflectance in the core alternates between light-colored sediments, rich in foraminiferal carbonate and silica, and darker sediments abundant in detrital organic carbon. These changes in the marine sediments are proxies for the prevailing atmospheric circulation regime, and the meridional position of the ITCZ in the region, which are both linked to the glacial-interglacial variability.

For the KS analysis, a 20-year moving average of the MD03-2621 record is calculated in order to align its resolution with that of the NGRIP record. Our analysis does identify the ``classical'' DO events, as seen in the NGRIP record \cite{rasmussen2014, bagniewski2021}. There is, however, no direct relationship in the identified longer-scale warm (grey bars) and cool intervals: For instance, some events appear to be merged in Fig.~\ref{MD03-2621}, e.g., GI 3 and 4, GI 9 and 10, GI 13 and 14, GI 15 and 16, and GI 22 and 23, while some events detected here, between 66 and 68~ka~BP, are not identified in NGRIP, and GI 2 is much longer than in NGRIP.

\begin{figure*}[!htbp]
\includegraphics[width=0.8\linewidth]{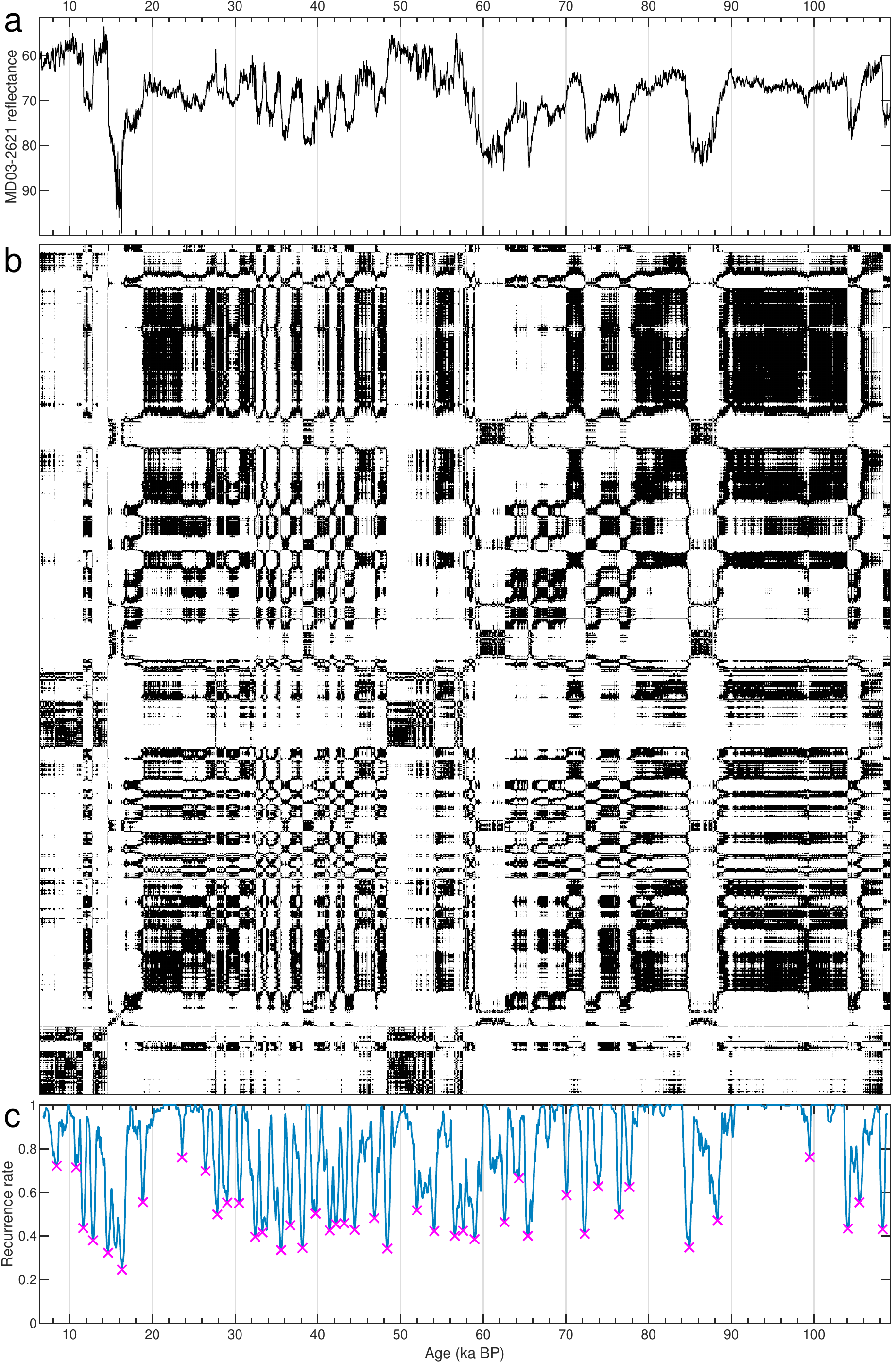}
    \caption{Recurrence Quantification Analysis (RQA)\cite{marwan2007,marwan2013} for the MD03-2621 reflectance record \cite{deplazes2013}. (a) Time series with vertical axis as in Fig.~\ref{MD03-2621}; (b) recurrence plot (RP); and (c) recurrence rate. Magenta crosses in panel (c) represent the transitions identified by the RQA.}
\label{Fig_RQA}
\end{figure*}

The RQA analysis \cite{marwan2007,marwan2013}, shown in Fig. \ref{Fig_RQA}, does identify the major transitions in the MD03-2621 reflectance record, including the relative significance of each. It does not, however, resolve smaller transitions that occur at the centennial time scale. Please see Bagniewski et al. \cite{bagniewski2021} for the explanation of the recurrence rate used to identify the transitions in the figure's panel (c).

\begin{figure*}[!htbp]
\includegraphics[width=0.8\linewidth]{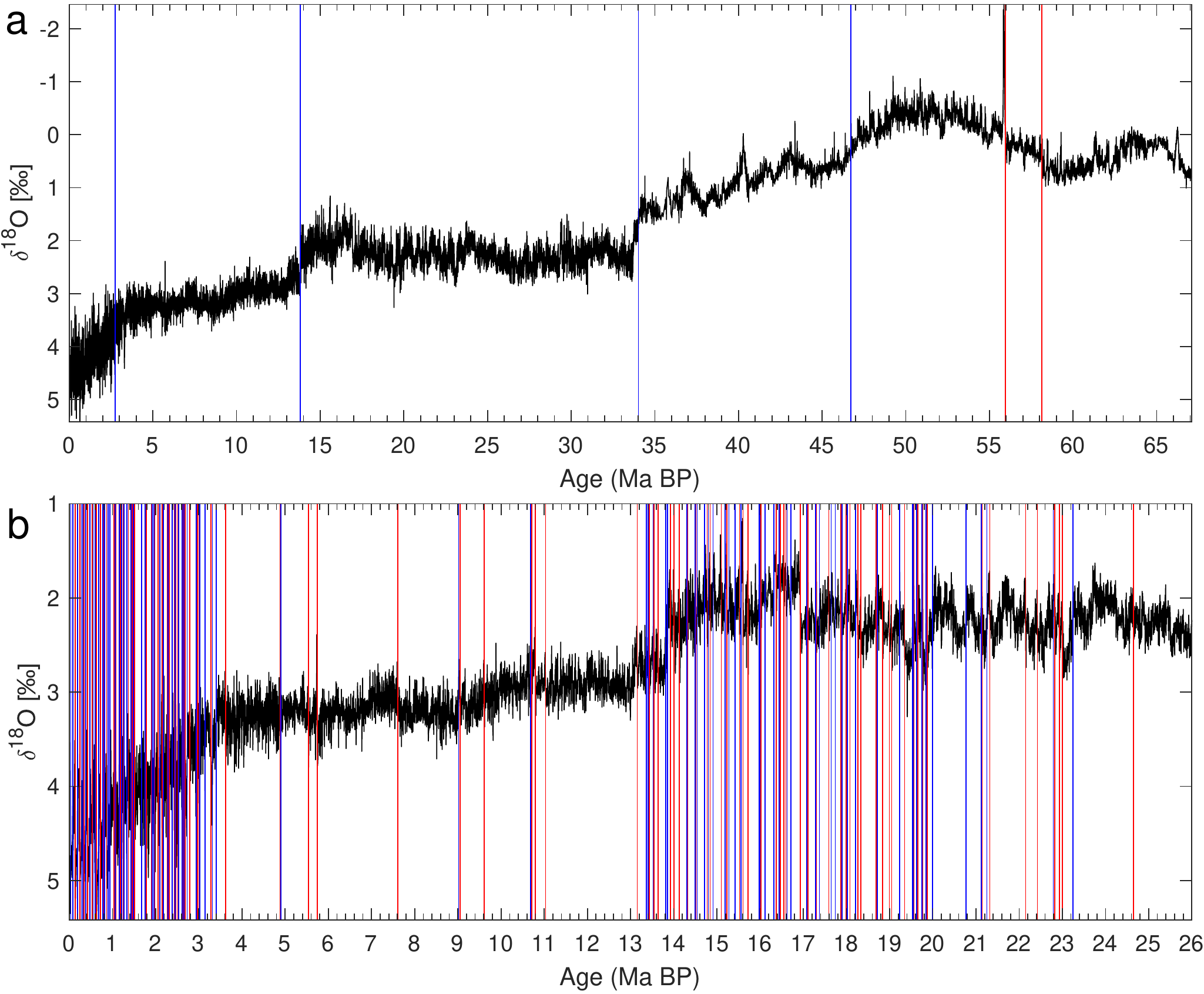}
\caption{CENOGRID stack of benthic $\delta^{18}$O \cite{westerhold2020}. Vertical lines represent
	transitions detected by the KS test \cite{bagniewski2021}, with red lines for warming transitions and blue lines for cooling ones, while grey bars represent warm episodes.
	(a) Transitions detected for the entire record using a fixed window length of $w = 5$~Myr, and
	(b) transitions detected for the interval 0--26~Ma~BP using $0.01 \le w \le 1.25$~Myr;
	see also the earlier subsection on ``Methodology.'' The vertical axes are reversed.}
\label{cenogr}
\end{figure*}

The CENOGRID stack of benthic $\delta^{18}$O \cite{westerhold2020} shown here in Fig.~\ref{cenogr} is a highly resolved 67~Myr composite from 14 marine records. Westerhold et al. \cite{westerhold2020} distinguished four climate states --- Hothouse, Warmhouse, Coolhouse, and Icehouse --- in this record, largely following changes in the polar ice volume. The composite in the figure \cite{westerhold2020} reconstructs in detail Earth's climate during the Cenozoic era, at a higher time resolution than the earlier compilation of Zachos and colleagues \cite{Zachos2001}.

Our KS analysis in Fig.~\ref{cenogr}a uses a window width $x \le w \le y$ and it identifies four major transitions towards higher $\delta^{18}$O values and two towards lower ones. The oldest threshold, at 58~Ma~BP, characterizes the transition between the moderately warm climate prevailing at the beginning of the Cenozoic to the hot conditions marked by the Early Eocene Climate Optimum between 54~Ma and 49-48~Ma~BP. The second transition corresponds to the short but intense warm event known as the PTEM, the Cenozoic's hottest one. The third transition marks the end of this hot interval and the return to the milder and relatively stable conditions that prevailed between 67~Ma and 58~Ma~BP.

The fourth threshold at 34~Ma is the Eocene--Oligocene Transition, the sharp boundary between the warm and the hot climatic conditions in the earlier Cenozoic and the Coolhouse and then Icehouse conditions prevailing later on\cite{Coxall.ea.2005}. This transition is an intriguing candidate for TP status in Earth's climate history \cite{Rousseau.ea.2022}. The fifth transition, at 14~Ma~BP, ends a rather stable climate interval between 34~Ma and 14~Ma, characterized by the build up of the East Antarctic ice sheet\cite{Miller.ea.1987, Flower.Kennett.1994}. This transition also marks the start of an increasing trend in benthic $\delta^{18}$O values \cite{Miller.ea.2020, Rohling.ea.2021}. The final transition marks the start of the Icehouse world and it is characterized by the emergence and development of the Northern Hemisphere ice sheets, with their further expansion occurring during the past 400~kyr; compare U1308 results in Fig.~\ref{ohrid} below.

Using a reduced window length on the last 26~Myr of the CENOGRID benthic record, many more abrupt transitions are detected in Fig.~\ref{cenogr}b. In particular, higher variability in the $\delta^{18}$O signal and much more frequent transitions are found during two intervals, namely 71 transitions over the last 3.5~Myr and 77 transitions between 13~Ma~BP and 20~Ma~BP. In contrast, the intervals 3.5--13~Ma~BP and 20--67~Ma~BP are characterized by a lower frequency of detected transitions, with 13 and 112 transitions, respectively. The former one of the two intervals with many jumps includes the Quaternary Period, which started 2.6 Myr ago and is well known for its high climate variability, due to the presence of large ice masses in the system \cite{Emiliani.Geiss.1959,Ghil.Chil.1987}.

The interval 13--20~Ma~BP, on the other hand, coincides with the exclusive use of the U1337 and U1338 records in constructing the CENOGRID stack, both of which are located in the eastern tropical Pacific. Higher sedimentation rates in these two cores might contribute to the higher variability $\delta^{18}$O observed in the stack record over this interval. For transitions detected with the shorter window for the entire CENOGRID stack, please see Fig.~S1 in the Supplementary Material.

\begin{figure*}[!htbp]
\includegraphics[width=0.8\linewidth]{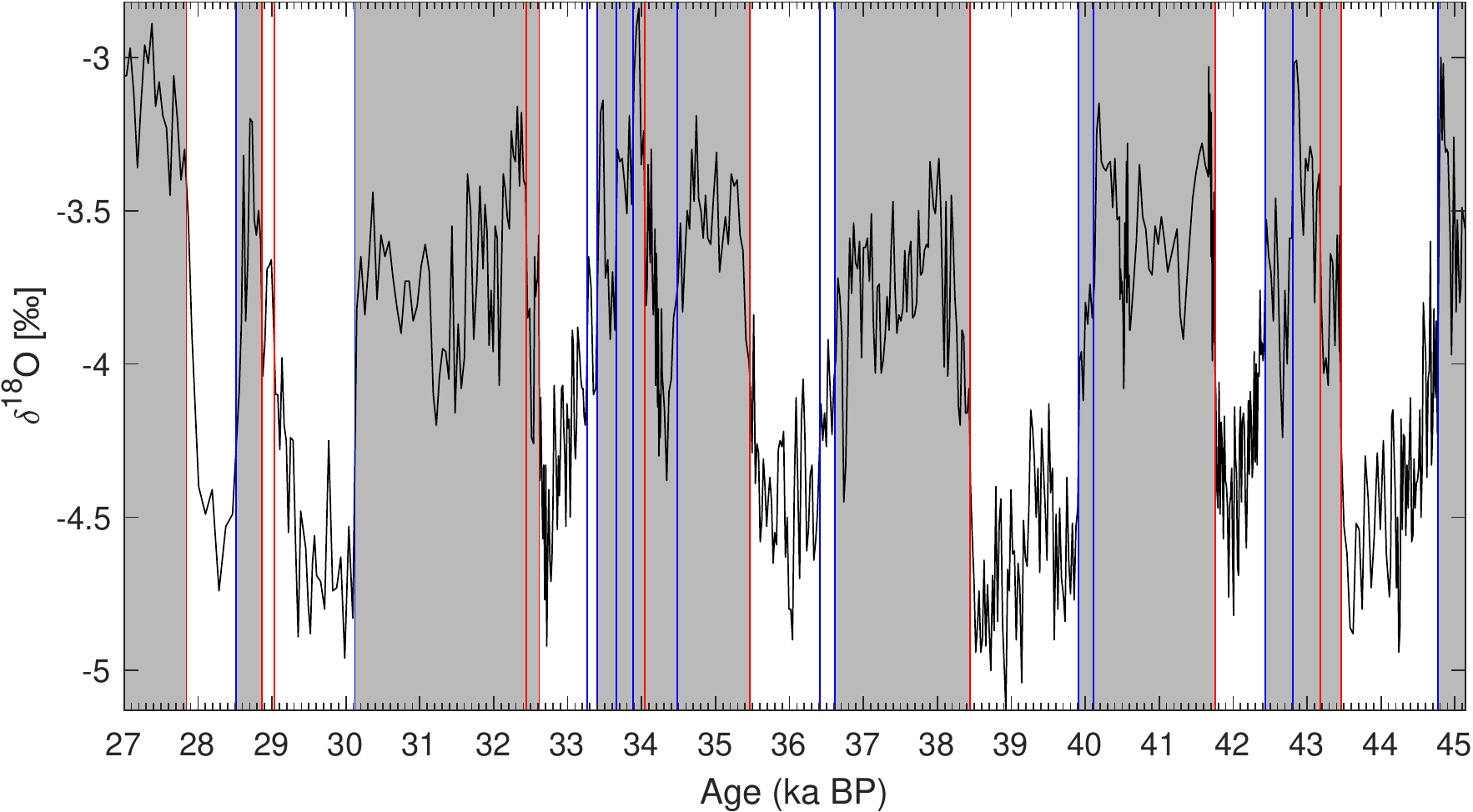}
    \caption{Paraiso Cave (PAR07) $\delta^{18}$O \cite{wang2017}. Vertical lines represent transitions detected by the KS algorithm \cite{bagniewski2021}, with red lines for a decrease in $\delta^{18}$O and blue lines for an increase in $\delta^{18}$O, while grey bars correspond to warm events.}
\label{paraiso}
\end{figure*}

The $\delta^{18}$O record from the Paraiso Cave, located in the Amazon rainforest, shows moisture patterns that undergo abrupt shifts during DO events; see Fig. \ref{paraiso}. It is likely that the Amazonian climate subsystem exhibits bistability, thus making it a potential tipping element, as changes in precipitation accelerate dieback of the forest\cite{Boers.ea.2017}. The Paraiso Cave record indicates that rainfall over the Amazon basin corresponds to global temperature changes, with dryer conditions during the last glacial period. This record exhibits negative correlation with the Chinese speleothem records \cite{cheng2016}, suggesting that rainfall in these two regions is in phase opposition.

\paragraph{Results for core intercomparisons.}

\begin{figure*}[!htbp]
\includegraphics[width=0.8\linewidth]{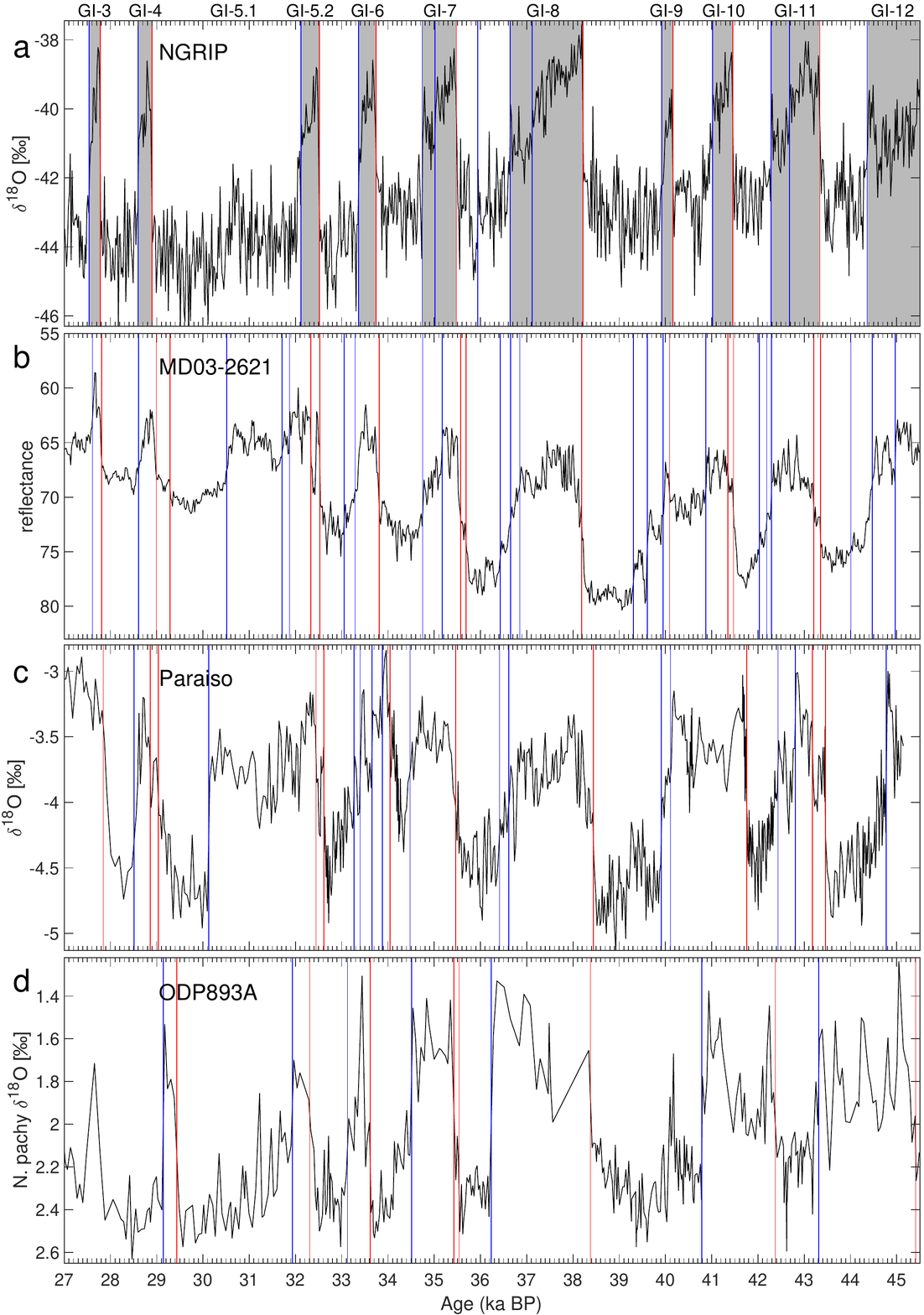}
    \caption{Comparison of four paleorecords over the 27 -- 45.5 ka BP interval: (a) NGRIP ice core $\delta^{18}$O \cite{rasmussen2014}; (b) MD03-2621 marine sediment reflectance \cite{deplazes2013}; (c) Paraiso Cave (PAR07) $\delta^{18}$O \cite{wang2017}; and (d) ODP893A marine sediment N. pachyderma $\delta^{18}$O \cite{hendy2002,hendy1999}. Vertical lines represent detected transitions, with red lines for warming transitions and blue lines for cooling ones. Grey bars in panel (a) represent interstadials identified for the NGRIP $\delta^{18}$O record, with top labels showing the Greenland Interstadials identified by Rasmussen et al. \cite{rasmussen2014}.}
\label{4cores}
\end{figure*}

In Fig.~\ref{4cores}, we compare four paleorecords of different types and from distinct locations: NGRIP $\delta^{18}$O, MD03-2621 reflectance, Paraiso Cave $\delta^{18}$O, and ODP893A planktic $\delta^{18}$O from the Santa Barbara Basin.\footnote{Note that a separate time axis appears in the figure for the NGRIP record, with the time unit being ``ka~b2k” rather than ``ka~BP.'' This is so because ice cores typically exhibit higher resolution than marine-sediment cores and 'b2k' refers to the year 2000 as the origin of past times \cite{rasmussen2014}, more precisely than the 'BP' of footnote~\ref{BP}.} Overall, the abrupt transitions in the four records appear to be fairly synchronous, with the main Greenland DO events from NGRIP also observed for the two marine records and the one speleothem record. The transitions that correspond to Greenland interstadials (GIs) GI-3 to GI-12 \cite{rasmussen2014} are identified in all four records, with only a few exceptions: GI-3 is not identified for the ODP893A record, due to an insufficient number of data points; and GI-5.1 is not identified in any record, except as a cooling event in MD03-2621 and in Paraiso cave. Also, GI-9 and GI-10 appear as a single interstadial in the Paraiso cave and ODP893A records, which is probably due to a decreased time resolution observed in both records during this time interval.

Finally, Fig.~\ref{ohrid} shows the comparison between the U1308 benthic $\delta^{18}$O record and the Lake Ohrid Total Inorganic Carbon (TIC) record over a 1.4~Myr time interval that includes multiple glacial-interglacial transitions.

The U1308 benthic $\delta^{18}$O record in panel (a) of the figure is a 3.1~Myr record located within the ice-rafted detritus belt of the North Atlantic \cite{ruddiman1977}. This record is a proxy for deep-water temperature and global ice volume, and it has enabled a detailed reconstruction of the history of orbital and millennial-scale climate variability during the Quaternary; it documents the changes in Northern Hemisphere ice sheets that follow the glacial--interglacial cycles \cite{hodell2016,rousseau2022}, as well as mode transitions identified at 2.55~Ma, 1.5~Ma, 1.25~Ma, 0.65~Ma and 0.35~Ma~BP.

The 1.25--0.65~Ma~BP interval in the figure corresponds to the Mid-Pleistocene Transition (MPT) interval with 1.25~Ma being followed by an increase in the amplitude of the glacial--interglacial fluctuations and a transition from a dominant periodicity near 40~kyr to one near 100~kyr; see Reichers et al. \cite{Riechers.ea.2022} and references therein. The most recent abrupt transition, at 0.35 Ma, leads to the strongest interglacial of the record, which corresponds to the marine isotope stage (MIS) 9. Here, our KS analysis agrees with the well established marine isotope stratigraphy of Lisiecki and Raymo \cite{lisiecki2005} by detecting rapid warmings that correspond to the classical terminations leading to interglacials, as well as rapid coolings that initiate glacial stages.

The Lake Ohrid TIC record in Fig.~\ref{ohrid}b shows glacial--interglacial variability in biomass over the past 1.4~Myr. Here, our KS analysis shows numerous abrupt transitions towards high TIC intervals that correspond to interglacial episodes, as well as matching ones of opposite sign. The interglacial ones are associated with forested environmental conditions that are consistent with odd-numbered MISs, as was the case for the U1308 record in the figure's panel (a). The glacial episodes, to the contrary, correspond to forestless conditions, that, according to the available timescale, are consistent with even-numbered MISs.

\begin{figure*}[!ht]
\includegraphics[width=1\linewidth]{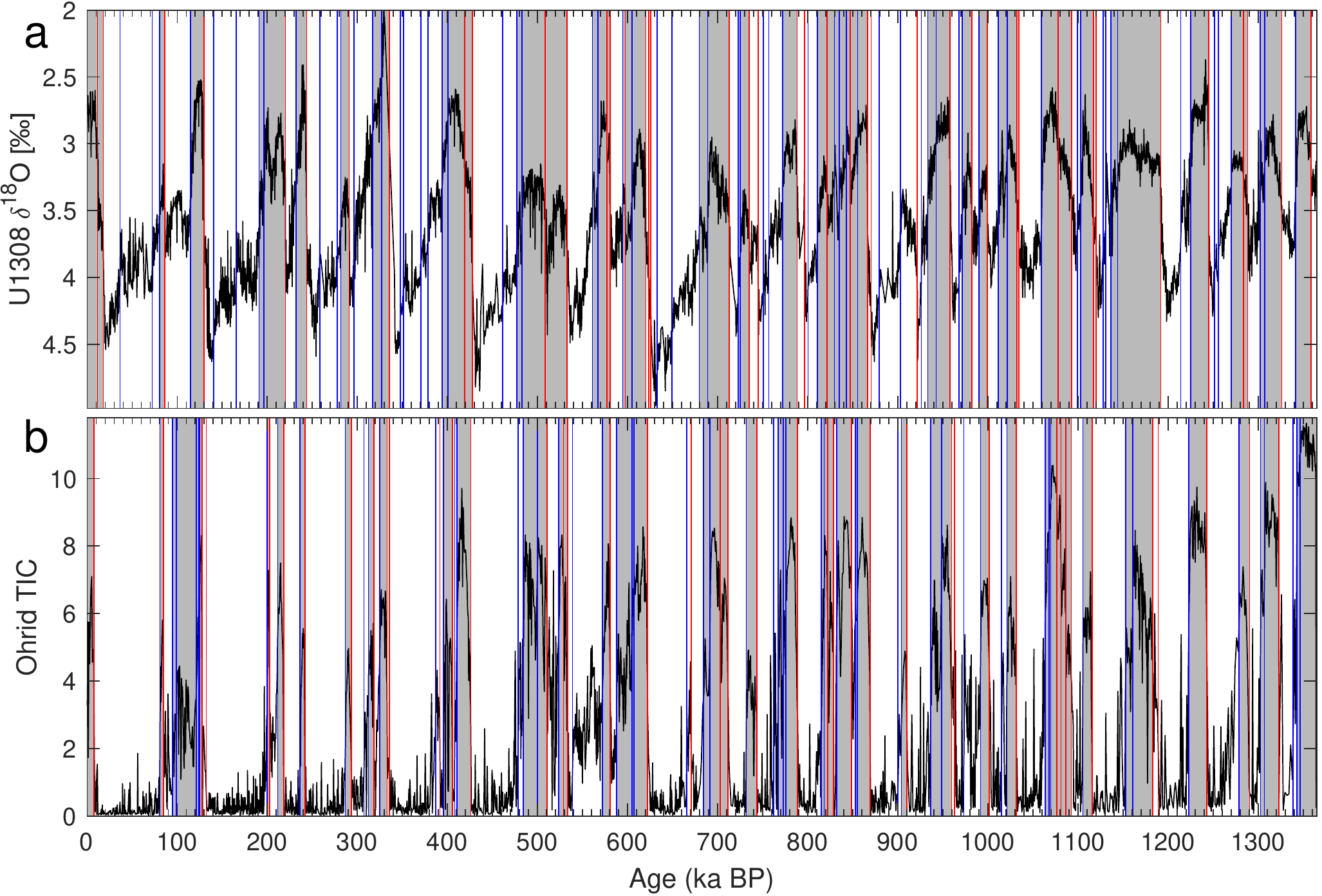}
\caption{Comparison of the benthic \emph{Cibicidoides sp.} $\delta^{18}$O record in
	(a) the U1308 marine-sediment core \cite{hodell2016}  and
	(b) Lake Ohrid TIC \cite{sadori2016}. Vertical lines represent transitions detected by our KS methodology \cite{bagniewski2021}, with red lines for warming transitions and blue lines for cooling ones, while grey bars correspond to interglacials (odd-numbered MISs) and white bars correspond to glacials (even-numbered MISs).}
\label{ohrid}
\end{figure*}

\section*{Discussion}

\paragraph{The two methods.}
The augmented KS test of Bagniewski et al. \cite{bagniewski2021} has detected abrupt transitions on different time scales in a variety of paleorecords. These include the NGRIP ice core, the Paraiso Cave speleothem, and the MD03-2621 and ODP893A marine sediment cores for the last climate cycle; the U1308 marine sediment core and Lake Ohrid TIC for the Middle and Late Pleistocene; and the CENOGRID marine sediment stack for the Cenozoic Era.
While possible mechanisms giving rise to the observed variability in these records have been discussed in previous studies, the objective, precise and robust dating of the main transitions using the augmented KS test may allow a more detailed and definitive analysis and modelling.

The transitions identified with the RQA methodology \cite{eckmann1987,marwan2007,marwan2013} do correspond to a subset of those found with the KS test, but several smaller-scale transitions identified by the latter have not been found with the former method. The RQA approach, though, does allow one to quantify the magnitude of each transition, and it may thus be useful for identifying key transitions.

\paragraph{Interpretation of findings.}
The high-resolution MD03-2621 reflectance record from the Cariaco Basin \cite{deplazes2013} in Fig.~\ref{MD03-2621} shows abrupt transitions that are largely in agreement with those detected by our KS test for the NGRIP record. Desplazes et al. \cite{deplazes2013} have argued that these transitions are driven by the ITCZ displacements that occurred primarily in response to Northern Hemisphere temperature variations. These authors indicated that the ITCZ migrated seasonally during mild stadials, but was permanently displaced south of the Cariaco Basin during the colder stadial conditions. The very high resolution of the MD03-2621 record allows a detailed comparison with the transitions identified for the NGRIP record. The fact that several small-scale NGRIP transitions are not identified by our KS test in the MD03-2621 record suggests that a TP linked to ITCZ migration was not crossed during these events. Furthermore, the KS test does reveal transitions that have not been recognized previously in either record \cite{rasmussen2014}, e.g., at 86.15~ka~BP, which we find in both the NGRIP and MD03-2621 records.

In the 67-Myr CENOGRID stack (Fig.~\ref{cenogr}) of benthic $\delta^{18}$O \cite{westerhold2020,Zachos2001}, we identified four major cooling transitions that culminate with the start of the Pleistocene, as well as two warming transitions, including the Paleocene-Eocene Thermal Maximum. These major jumps agree with those identified in Westerhold et al. \cite{westerhold2020}. However, using a shorter window length, many more transitions were detected during the Pleistocene, well known for its higher climate variability \cite{Emiliani.Geiss.1959, Ghil.Chil.1987}, as well as during the early Miocene, i.e., between the mid-Miocene transition at 14~Ma~BP \cite{Flower.Kennett.1994} and the Oligocene-Miocene transition at 23~Ma~BP \cite{miller1991,boulila2011,shackleton2000b}. A possible reason is the fact that the 13--20~Ma~BP interval in the stack has been constructed using records from the eastern Tropical Pacific. This region has been characterized by high upwelling rates and sedimentation rates in the past \cite{pinero2013}. While the CENOGRID composite has a uniform resolution, higher sedimentation rates could affect the resolution of the original records and, therewith, even the variability seen in lower-resolution sampling. The fact that the source region of the data has a large effect on the frequency of detected transitions implies that caution is needed when using individual records as proxies for global climate.

The Paraiso Cave record \cite{wang2017} from the eastern Amazon lowlands in Fig.~\ref{paraiso} shows dryer conditions during the last glacial period, with abrupt transitions matching those that correspond to the DO events identified in the NGRIP ice core record. The Amazon record is also in fairly good agreement with the nearby MD03-2621 marine sediment record. Notably, several of NGRIP's DO events appear combined in both records, namely GI-5.1 and GI-5.2, as well as GI-9 and GI-10, which might indicate that a climate change event over Greenland did not trigger a tipping event in the Amazon basin. Alternatively, the first merging may question the separation of the classical GI-5 event \cite{dansgaard1993} into two distinct events, 5.1 and 5.2.

The comparison of four records on the same time scale in Fig. \ref{4cores} demonstrates that a signal of abrupt climate change is detected when using the KS method with similar accuracy for different types of paleodata. The differences in the dates of the transitions may be largely explained by the use of different age models in each of the records, with MD03-2621 fine-tuned to the NGRIP chronology, and ODP893A fine-tuned to the GISP2 chronology, while the NGRIP and Paraiso Cave records were independently dated. Notably, ODP893A data prior to GI-8 appear misaligned with the other three records. The chronology of jumps in ODP893A indicates that the warm interval between 40.8~ka~BP and 42.4~ka~BP may correspond to an event spanning GI-9 and GI-10, while the warm interval between 43.3~ka~BP and 45.3~ka~BP may correspond to GI-11.

In addition to the classical GI transitions, several additional jumps are identified in each of the marine and cave records. These jumps may be the representative of local climate changes or, in some cases, be artifacts of sampling resolution or measurement error. Stronger evidence for a local or regional event may be obtained when comparing two nearby records, like the Paraiso Cave in the Amazon and the MD03-2621 marine record from the Cariaco Basin. In both records, the start of GI-5.2 is represented as two warming transitions, in contrast with the NGRIP and ODP893A records, where only one sharp transition is present. Likewise, the end of GI-9 in the two tropical, South American records appears as two successive cooling events. These results point to the potential of using the KS method to improve fine-tuning the synchroneity of two or more records, when an independent dating method is not available.

The comparison in Fig.~\ref{ohrid} of KS-detected transitions in marine core U1308 \cite{hodell2016} and in Lake Ohrid TIC \cite{sadori2016} identifies the transitions between individual glacials and interglacials. In the two records, the transitions are well identified, despite the resolution of the two being different, and they offer a precise dating for the chronology of the past glacial cycles. The transitions are overall in good agreement between the two records. Still, the warm events in Lake Ohrid are essentially atmospheric and thus have often a shorter duration than in the marine record, while cooling transitions precede those in the deep ocean by several thousand years. This could indicate that a significant time lag is present, either in ice sheet growth in response to atmospheric cooling, or in the propagation of the cooling signal into the deep North Atlantic.

Furthermore, there are atmospheric interglacial episodes missing in the oceanic U1308 record. This mismatch between the lake record  \cite{sadori2016} and the marine one \cite{hodell2016} could be attributed to the fact that the KS test does sometimes find more transitions in one record than in another one, even for the same type of proxy and within the same region. Such occurrences may be due to the local environment, the sampling method, or some aspect of the KS method itself. When this is the case, it is important --- although not always feasible --- to find one or more additional records covering a similar time interval with a similar resolution, in order to shed further light on the mismatch between the two original records.

\section*{Concluding remarks}

The carefully selected, high-quality paleoproxy records in the PaleoJump database \cite{BRG_PJ}, \href{https://paleojump.github.io}{https://paleojump.github.io},  have different temporal scales and a global spatial coverage; see again Fig. \ref{map}. These records provide an easily accessible resource for research on potential tipping elements in Earth’s climate. Still, major gaps in the marine sediment records exist in the Southern Hemisphere, especially in the Indian and Pacific Oceans. Only sparse terrestrial data are available from the high latitudes in both hemispheres, due to the recent glaciation. The only data from the African continent come from two East African lake sediment records. Even though much information is at hand from the more than 100 sites listed in this paper, more records are needed to fill geographical and temporal gaps, especially in the Southern Hemisphere.

The examples given in the paper on abrupt-transition identification demonstrate the usefulness of the records included in PaleoJump for learning about potential tipping events in Earth's history and for comparing such events across different locations around the world. The accessibility of such high-quality records is an invaluable resource for the climate modelling community that requires comparing their results across a hierarchy of models \cite{Ghil.2019, Schneider.Dick.1974} with observations.

We also demonstrated the usefulness of the KS test \cite{bagniewski2021} for establishing precisely the chronology of Earth's main climatic events. The newly developed tool for automatic detection of abrupt transitions may be applied to different types of paleorecords, allowing to objectively and robustly characterize the tipping phenomenon for climate subsystems already suspected of being subject to tipping \cite{lenton2008}, but also to identify previously unrecognized tipping elements in past climates. The observational descriptions of tipping that can be obtained from PaleoJump using our KS methodology, combined with the application of Earth System Models, can help improve the understanding of the bifurcation mechanisms of global and regional climate and identify possible TPs for future climates.

Our results also indicate that paleorecord interpretations may vary, since the abrupt transitions identified in them will depend on the time scale and type of variability that is investigated. For example, the KS method's parameters \cite{bagniewski2021} may be changed when studying different proxy record resolutions, affecting the frequency and exact timing of the TPs that are identified.

The agreement in timing and pattern between jumps in distinct records can confirm the correctness of each record separately, as well as of the inferences on climate variability drawn from these jumps.
Specifically, the ability of the KS method to identify matching small-scale transitions in different high-resolution records may be used to validate these transitions as being the result of genuine global or regional climatic events, as opposed to just sampling errors. Furthermore, significant differences in records that are, overall, in good agreement with each other may help decode the chronology of tipping events or an approximate range for a tipping threshold. A fortiori, the differences in timing and pattern between jumps in distinct records that we also found emphasize the importance of a consistent dating methodology.

The broad spatial coverage of the PaleoJump database \cite{BRG_PJ}, \href{https://paleojump.github.io}{https://paleojump.github.io}, with its records that vary in their nature --- ice, marine and land --- as well as in their length and resolution, will facilitate research on tipping elements in Earth’s climate, including the polar ice sheets, the Atlantic Meridional Overturning Circulation, and the tropical rainforests and monsoon systems. Furthermore, it will support establishing improved criteria on where and how to collect data for reliable early warning signals of impending TPs.

\bibliography{ms}

\section*{Acknowledgements}

This paper is Tipping Points in the Earth System (TiPES) contribution \#XX. This study has been funded by the European Union's Horizon 2020 research and innovation programme (grant agreement No. 820970). This is LDEO contribution.

\section*{Author contributions statement}

W.B., M.G. and D.D.R. outlined the manuscript, W.B. carried out the computations and drew the plots, and W.B., M.G. and D.D.R. analyzed the results. W.B. drafted the manuscript and all authors reviewed and finalized it.

\section*{Competing interests}

The authors declare no competing interests.

\end{document}


\flushbottom
\maketitle
%
%
\thispagestyle{empty}

\newpage


\section*{\centering\LARGE{Supplementary Tables}}
\renewcommand{\tablename}{Supplementary Table S}

The sites of proxy records included in the PaleoJump database have been compiled in five tables according to their geological nature. For each site, the available data have been analyzed to determine the essential information, which is given in the tables below: location, depth/elevation, temporal range, maximum temporal resolution, and types of paleoproxies. This information is accompanied by links to the original data and the associated publications. The ``maximum resolution'' value in the tables is calculated as the maximum of the average temporal resolution for a 10-kyr time interval, excluding the Holocene and the late deglacial, i.e. the last 14~000 years, during which the proxy time resolution is frequently much higher than for the older part of the record. This was done so as to allow a more accurate comparison of the centennial- and millennial-scale variabilities between different records of the glacial and earlier interglacials periods.

\section*{1. Records included in the PaleoJump database}

\begin{longtable}{@{}lllllll@{}}
\toprule
\textbf{Site name}  & \textbf{Location} & \textbf{Depth} & \textbf{Age} & \textbf{Res.} & \textbf{Proxies}                                             \\* \midrule
\endhead
\bottomrule
\endfoot
\endlastfoot
MD95-2010 \cite{dokken1999}  & 66.684, 4.566 & 1226 m & 67 - 10 ka & 35 y  & pla $\delta^{18}$O; ben $\delta^{18}$O; pla $\delta^{13}$C; ben $\delta^{13}$C; IRD\\
ODP162-983 \cite{barker2019} & 60.403, -23.641& 1984 m & 805 - 0 ka  & 97 y  & IRD; \%NPS\\
SO82-5 \cite{vankreveld2000,waelbroeck2019} & 59.186, -30.905 & 1394 m & 57 - 15 ka & 62 y & pla $\delta^{18}$O; ben $\delta^{18}$O; pla $\delta^{13}$C; ben $\delta^{13}$C; SST; IRD \\
MD95-2006 \cite{dickson2008}& 57.03, -10.058& 2122 m  & 56 - 40 ka  & 91 y  & pla $\delta^{18}$O; ben $\delta^{18}$O; pla $\delta^{13}$C; ben $\delta^{13}$C; SST; IRD  \\
JPC-13 \cite{hodell2010}& 53.057, -33.53& 3082 m  & 128 - 7 ka  & 71 y  & pla $\delta^{18}$O; pla $\delta^{13}$C; ben $\delta^{18}$O; ben $\delta^{13}$C; lithics  \\
U1308 \cite{hodell2016,hodell2008} & 49.878, -24.238 & 3871 m  & 3143 - 0 ka & 118 y & ben $\delta^{18}$O; ben $\delta^{13}$C; bulk carbonate $\delta^{18}$O; Ca/Sr; \\ & & & & & Si/Sr; pla $\delta^{18}$O  \\
MD01-2412  \cite{harada2006}  & 44.523, 145.003 & 1225 m  & 116 - 0 ka  & 113 y & SST \\
MD99-2331 \cite{eynaud2009,davtian2021,naughton2009,sanchez2017}  & 42.15, -9.683  & 2120 m  & 160 - 16 ka & 137 y & pla $\delta^{18}$O; IRD; SST; pollen; temperate forest pollen \\
U1313 \cite{bolton2018,naafs2020,naafs2013} & 41, -32.957 & 3426 m  & 4.3 - 0 Ma & 185 y & pla $\delta^{18}$O; ben $\delta^{18}$O; SST; Qz/Cal\\
MD95-2040 \cite{voelker2010,deabreu2003} & 40.582, -9.861 & 2465 m & 360 - 0 ka & 78 y & pla $\delta^{18}$O; pla $\delta^{13}$C; ben $\delta^{18}$O; ben $\delta^{13}$C; SST; IRD  \\
MD95-2039  \cite{eynaud2009,salgueiro2014}  & 40.579, -10.349& 3381 m  & 51 - 0 ka & 107 y & pla $\delta^{18}$O; IRD; SST\\
MD01-2443 \cite{hodell2013,voelker2010} & 37.881, -10.176  & 2925 m  & 433 - 86 ka & 188 y & pla $\delta^{18}$O; pla $\delta^{13}$C; ben $\delta^{18}$O; SST\\
MD95-2042 \cite{eynaud2009,martrat2007,shackleton2000,davtian2021} & 37.8, -10.167 & 3146 m & 418 - 0 ka & 83 y & pla $\delta^{18}$O; ben $\delta^{13}$C; pla $\delta^{13}$C; ben $\delta^{18}$O; SST\\
U1385 \cite{hodell2015,bahr2018} & 37.571, -10.126 & 2587 m & 1.4 - 0 Ma & 47 y & pla $\delta^{18}$O; reflectance; ben $\delta^{18}$O; SST\\
MD01-2444  \cite{martrat2007,hodell2013}  & 37.565, -10.134 & 2656 m  & 420 - 0 ka  & 86 y  & pla $\delta^{18}$O; ben $\delta^{18}$O; reflectance; SST  \\
MD99-2341 \cite{eynaud2009} & 36.389, -7.066  & 582 m & 49 - 1 ka & 104 y & pla $\delta^{18}$O \\
ODP977a \cite{martrat2007}& 36.032, -1.955& 1984 m  & 244 - 0 ka  & 181 y & SST \\
MD99-2339 \cite{voelker2006,salgueiro2014}  & 35.886, -7.528  & 1177 m  & 47 - 0 ka  & 38 y  & pla $\delta^{18}$O; ben $\delta^{18}$O; pla $\delta^{13}$C; ben $\delta^{13}$C; SST; \\ & & & & & grain size \\
M40/4\_SL71 \cite{beuscher2017,ehrmann2021,weldeab2003}& 34.811, 23.194 & 2788 m  & 182 - 0 ka  & & XRF; Clay; pla $\delta^{18}$O; \\
ODP893A  \cite{hendy2002,hendy1999,hendy2003}& 34.28, -120.03 & 576 m & 65 - 0 ka & 41 y  & pla $\delta^{18}$O; pla $\delta^{13}$C; ben $\delta^{18}$O \\
U1429 \cite{clemens2018}  & 31.617, 128.998 & 732 m & 393 - 0 ka  & 75 y  & pla $\delta^{18}$O; ben $\delta^{18}$O; ben $\delta^{13}$C; pla $\delta^{13}$C; SST \\
MD02-2575 \cite{nurnberg2008} & 29.002, -87.119 & 847 m & 400 - 1 ka  & 135 y & pla $\delta^{18}$O; ben $\delta^{18}$O; SST \\
MD04-2876 \cite{pichevin2007} & 24.843, 64.008& 828 m & 50 - 0 ka & 169 y & $\delta^{15}$N; total N; TOC\\
SO90-93KL  \cite{schulz1998}  & 23.583, 64.217 & 1802 m  & 109 - 1 ka  & 177 y & pla $\delta^{18}$O \\
SO130-289KL \cite{deplazes2013,deplazes2014} & 23.122, 66.497  & 571 m & 79 - 2 ka & 0.2 y & reflectance; grain size; TOC; ... \\
SO90-136KL \cite{schulz1998}  & 23.117, 66.5& 568 m & 66 - 2 ka & 63 y  & TOC  \\
SO90-111KL \cite{schulz1998}& 23.1, 66.483& 775 m & 62 - 2 ka & 72 y  & TOC\\
ODP658C \cite{zhao1995}& 20.75, -18.583 & 2263 m  & 84 - 0 ka & 135 y & SST\\
SO188-17286-1 \cite{lauterbach2020}& 19.743, 89.879  & 1428 m  & 129 - 0 ka  & 158 y & pla $\delta^{18}$O; ben $\delta^{18}$O; SST  \\
U1446 \cite{clemens2021}& 19.083, 85.733 & 1440 m  & 1.46 - 0 Ma & & Rb/Ca \\
GeoB9526-5 \cite{zarriess2010}& 12.435, -18.057 & 3223 m  & 72 - 1 ka & 145 y & Fe/K; SST; ben $\delta^{18}$O \\
NIOP905 \cite{jung2009}& 10.767, 51.951 & 1580 m  & 88 - 1 ka & 145 y & ben $\delta^{18}$O; ben $\delta^{13}$C \\
MD03-2621 \cite{deplazes2013} & 10.678, -64.972& 847 m & 109 - 6 ka  & 0.1 y & reflectance  \\
MD97-2141 \cite{rosenthal2003} & 8.78, 121.28 & 3633 m  & 395 - 5 ka  & 51 y  & pla $\delta^{18}$O \\
MD98-2181 \cite{stott2002,saikku2009} & 6.3, 125.83& 2114 m  & 68 - 0 ka & 66 y  & pla $\delta^{18}$O; pla $\delta^{13}$C; ben $\delta^{18}$O; SST \\
MD03-2707 \cite{weldeab2007} & 2.502, 9.395 & 1295 m  & 155 - 0 ka PB  & 92 y  & pla $\delta^{18}$O; SST\\
TR163-22 \cite{lea2006}& 0.52, -92.4& 2830 m  & 135 - 1 ka  & 169 y & pla $\delta^{18}$O; ben $\delta^{18}$O; SST\\
SO189-039KL \cite{mohtadi2014}& -0.79, 99.908 & 517 m  & 45 - 0 ka  & & pla $\delta^{18}$O; SST\\
GeoB6518-1 \cite{weijers2009,rampen2012}& -5.588, 11.222 & 962 m  & 43 - 0 ka  & & pla $\delta^{18}$O; ben $\delta^{18}$O; TOC; SST\\
GeoB7139-2 \cite{depolholz2007} & -30.2, -71.983  & 3267 m  & 70 - 1 ka & 53 y  & $\delta^{15}$N; total N\\
MD03-2611G \cite{dedeckker2020}  & -36.73, 136.548& 2420 m  & 94 - 0 ka & 15 y  & pla $\delta^{18}$O; pla $\delta^{13}$C; SST; quartz; Ti& \\
ODP1089 \cite{hodell2003}& -40.93, 9.9& 4621 m  & 615 - 0 ka  & 189 y & pla $\delta^{18}$O; ben $\delta^{18}$O; pla $\delta^{13}$C; ben $\delta^{13}$C \\
TNO57-21 \cite{barker2014}& -41.1, 7.8 & 4981 m  & 99 - 0 ka & 112 y & pla $\delta^{18}$O; \%NPS \\
MD02-2588 \cite{ziegler2013} & -41.332, 25.828 & 2907 m  & 353 - 0 ka  & 246 y & ben $\delta^{18}$O; ben $\delta^{13}$C; pla $\delta^{13}$C \\
ODP181-1123 \cite{elderfield2012} & -41.786, -171.499 & 3290 m  & 1.5 - 0 Ma  & 388 y & ben $\delta^{18}$O; ben $\delta^{13}$C; Mg/Ca \\
MD07-3076Q \cite{gottschalk2015,riveiros2010}  & -44.153, -14.228  & 3770 m  & 67 - 1 ka & 61 y  & pla $\delta^{18}$O; pla $\delta^{13}$C; SST\\
MD97-2120 \cite{pahnke2003} & -45.534, 174.931 & 1210 m  & 151 - 4 ka  & 103 y & pla $\delta^{18}$O; SST \\
MD88-770 \cite{rickaby1999}& -46.017, 96.467 & 3290 m  & 149 - 5 ka  & 187 y & SST\\
CENOGRID \cite{westerhold2020}& N/A & N/A& 67.1 - 0 Ma & 2000 y& ben $\delta^{18}$O; ben $\delta^{13}$C \\* \bottomrule
\caption{Marine Sediment Cores, ordered by latitude. Res.: temporal resolution, pla: planktic, ben: benthic, SST: sea surface temperature, IRD: ice rafted detritus, TOC: total organic carbon.}
\label{tab:table1}\\
\end{longtable}

\begin{longtable}{@{}lllllll@{}}
\toprule
\textbf{Site name}  & \textbf{Location} & \textbf{Elevation} & \textbf{Age} & \textbf{Res.} & \textbf{Proxies} \\* \midrule
\endhead
\bottomrule
\endfoot
\endlastfoot
NEEM \cite{rasmussen2013}& 77.45, -51.06 & 2545 m  & 108 - 0 ka  & 10 y & $\delta^{18}$O; Ca$^{2+}$; Na+; ...\\
NGRIP \cite{rasmussen2014,baumgartner2014} & 75.1, -42.32& 2925 m  & 122 - 0 ka  & 20 y & $\delta^{18}$O; Ca$^{2+}$; dust; CH4\\
GRIP \cite{rasmussen2014}& 72.58, -37.63 & 3200 m  & 104 - 0 ka  & 20 y & $\delta^{18}$O; Ca; ...  \\
Guliya \cite{thompson1997}  & 35.28, 81.48& 6200 m  & 132 - 0 ka & 400 y  & $\delta^{18}$O; dust; ...\\
TALDICE \cite{bazin2013,vallelonga2013} & -72.783, 159.067 & 2315 m  & 314 - 0 ka  & 39 y & $\delta^{18}$O; CH4; Fe \\
EPICA EDML \cite{barbante2006} & -75.003, 0.068  & 2416 m& 150 - 0 ka  & 19 y & $\delta^{18}$O \\
EPICA Dome C \cite{jouzel2007,luthi2008,loulergue2008,lambert2008,extier2018,lambert2012} & -75.1, 123.35  & 3189 m  & 802 - 0 ka  & 39 y & $\delta$D; delta T; CH4; CO2; dust; $\delta^{18}$O; Ca$^{2+}$  \\
Dome Fuji \cite{uemura2018}  & -77.32, 38.7 & 3810 m  & 716 - 0 ka & 34 y & $\delta^{18}$O; dust \\
Vostok \cite{petit1999}  & -78.47, 106.8  & 3488 m  & 423 - 0 ka & 55 y  & $\delta$D, delta T, CH4, dust, ...\\
WAIS Divide \cite{wais2015}  & -79.468, -112.087& 1806 m  & 68 - 0 ka & 12 y & $\delta^{18}$O; ... \\
Synth. Greenland \cite{barker2011} & N/A& 2135 m  & 798 - 5 ka  & 50 y & $\delta^{18}$O \\* \bottomrule
\caption{Ice cores, ordered by latitude.}
\label{tab:table2}\\
\end{longtable}

\begin{longtable}{@{}lllllll@{}}
\toprule
\textbf{Site name}  & \textbf{Location} & \textbf{Elevation} & \textbf{Age} & \textbf{Res.} & \textbf{Proxies}\\* \midrule
\endhead
\bottomrule
\endfoot
\endlastfoot
Gassel Tropfsteinhöhle Cave \cite{moseley2020} & 47.823, 13.843  & 1225 m  & 108 - 77 ka& 5 y& $\delta^{18}$O\\
Grete-Ruth Cave \cite{moseley2020} & 47.543, 12.027  & 1435 m  & 111 - 103 ka & 12 y & $\delta^{18}$O\\
Hölloch im Mahdtal Cave \cite{moseley2020}& 47.378, 10.151  & 1438 m  & 74.4 - 73.6 ka & 5 y& $\delta^{18}$O\\
Schneckenloch Cave \cite{moseley2020,boch2011} & 47.375, 10.068 & 1285 m  & 118 - 64 ka& 7 y& $\delta^{18}$O\\
Grosser Baschg Cave \cite{moseley2020,boch2011}  & 47.25, 9.667 & 785 m & 87 - 81 ka & 14 y & $\delta^{18}$O\\
Villars Cave (Vil-stm09) \cite{genty2003,genty2010} & 45.442, 0.785 & 175 m & 83 - 31 ka & 88 y & $\delta^{18}$O; $\delta^{13}$C\\
Kesang Cave \cite{cheng2012} & 42.867, 81.75& 2000 m  & 500 - 52 ka& 23 y & $\delta^{18}$O\\
Sofular Cave \cite{fleitmann2009}& 41.416, 31.934& 700 m & 50 - 0 ka& 18 y & $\delta^{18}$O; $\delta^{13}$C\\
Leviathan Cave \cite{lachniet2014}  & 37.831, -115.607& 2400 m  & 174 - 0 ka & 50 y & $\delta^{18}$O\\
Dim Cave (Dim-E3) \cite{unal2015}& 36.54, 32.11& 232 m & 90 - 13 ka & & $\delta^{18}$O; $\delta^{13}$C\\
Devils Hole \cite{moseley2016} & 36.425, -116.291& 719 m & 204 - 5 ka & 65 y & $\delta^{18}$O; $\delta^{13}$C\\
Fort Stanton Cave \cite{asmerom2010}& 33.507, -105.494 & 1864 m  & 56 - 11 ka & 23 y & $\delta^{18}$O\\
Hulu Cave (MSD, MSL) \cite{wang2001}  & 32.5, 119.16& 86 m& 76 -18 ka& 68 y & $\delta^{18}$O; U/Th ages\\
Cave of the Bells \cite{wagner2010}& 31.729, -110.768& 1639 m  & 53 - 11 ka & 18 y & $\delta^{18}$O\\
Sanbao Cave \cite{wang2008,cheng2016} & 31.667, 110.433 & 1900 m  & 641 - 0 ka & & $\delta^{18}$O; U/Th ages  \\
Soreq Cave \cite{bar2003} & 31.45, 35.03 & 400 m  & 184 - 0 ka & 44 y & $\delta^{18}$O; $\delta^{13}$C\\
Bittoo Cave \cite{kathayat2016} & 30.79, 77.776  & 3000 m  & 284 - 0 ka & 18 y & $\delta^{18}$O; $\delta^{13}$C\\
Abaco Island Cave \cite{arienzo2015}& 26.23, -77.16 & -17 m & 64 - 14 ka & 14 y & $\delta^{18}$O; $\delta^{13}$C; Sr/Ca; Mg/Ca\\
Dongge Cave \cite{kelly2006} & 25.28, 108.08 & 680 m & 146 - 99 ka& 20 y & $\delta^{18}$O; U/Th ages\\
Moomi Cave \cite{burns2003}& 12.5, 54& 400 m & 53 - 40 ka & 15 y & $\delta^{18}$O\\
Terciopelo Cave \cite{lachniet2009} & 10.167, -85.333 & 370 m & 98 - 24 ka & 100 y  & $\delta^{18}$O\\
Secret Cave \cite{carolin2013} & 4.085, 114.85  & 250 m & 100 - 32 ka& 92 y & $\delta^{18}$O\\
Santiago Cave \cite{mosblech2012} & -3.017, -78.133  & 980 m & 94 - 6 ka & & $\delta^{18}$O\\
Paraiso (PAR07) \cite{wang2017}  & -4.067, -55.45& 60 m& 45 - 18 ka & 21 y & $\delta^{18}$O; $\delta^{13}$C\\
Pacupahuain Cave \cite{kanner2012}& -11.24, -75.82 & 3800 m  & 50 - 16 ka & 27 y & $\delta^{18}$O; $\delta^{13}$C\\
Ball Gown Cave \cite{denniston2013}& -17.33, 124.083 & 100 m  & 40 - 8 ka & & $\delta^{18}$O\\
Botuvera Cave \cite{cruz2005} & -27.22, -49.16& 230 m & 116 - 0 ka & 95 y & $\delta^{18}$O; $\delta^{13}$C\\
Hollywood Cave \cite{whittaker2011}  & -41.95, 171.467  & 130 m & 73 - 11 ka & 58 y & $\delta^{18}$O; $\delta^{13}$C \\
China cave composite \cite{cheng2016} & N/A& N/A & 641 - 0 ka & 22 y & $\delta^{18}$O; U/Th ages
\\* \bottomrule
\caption{Speleothems, ordered by latitude.}
\label{tab:table3}\\
\end{longtable}

\begin{longtable}{@{}lllllll@{}}
\toprule
\textbf{Site name}  & \textbf{Location} & \textbf{Elevation} & \textbf{Age} & \textbf{Res.} & \textbf{Proxies} \\* \midrule
\endhead
\bottomrule
\endfoot
\endlastfoot
ELSA-Eifel loess stack \cite{seelos2009} & 50.16, 6.83 & 402.5 m & 132 - 0 ka & 100 y  & dust \\
Nussloch loess \cite{rousseau2017,moine2017}  & 49.316, 8.722 & 180 m  & 130 - 18 ka  &  & grain size, $\delta^{13}$C, snails, earth worms,\\ & & & & & magnetic susc., paleosol-loess  \\
Dunaszekcso loess \cite{ujvari2014} & 46.09, 18.763  & 135 m & 150 - 20 ka  & & grain size, paleosol-loess \\
Gulang loess \cite{sun2012}& 37.49, 102.88 & 2400 m  & 60 - 0 ka& 50 y & grain size \\
Jingyuan loess \cite{sun2010,sun2012}  & 36.35, 104.6& 2210 m  & 80 - 0 ka& 26 y & grain size \\
Jiyuan loess \cite{yang2014}& 37.14, 107.39 & 1730 m  & 130 - 0 ka & & grain size \\
Zichang loess \cite{yang2014} & 37.14, 109.85 & 1265 m  & 249 - 0 ka & & grain size \\
Hongde loess \cite{yang2014}& 36.77, 107.21 & 1640 m  & 249 - 0 ka & & grain size \\
Huanxian loess \cite{yang2014}  & 36.65, 107.26 & 1500 m  & 249 - 0 ka & & grain size \\
Huachi loess \cite{yang2014}& 36.34, 107.93 & 1395 m  & 249 - 0 ka & & grain size\\
Xinzhuangyuan loess \cite{yang2014} & 36.19, 104.73 & 2110 m  & 220 - 0 ka & & grain size \\
Lijiayuan loess \cite{yang2014} & 36.12, 104.86 & 1850 m  & 220 - 0 ka & & grain size\\
Yimaguan loess \cite{hao2012} & 35.917, 107.617  & 1500 m  & 879 - 0 ka & 197 y & grain size; magnetic susceptibility \\
Luochuan loess \cite{hao2012} & 35.717, 109.417 & 1100 m  & 884 - 0 ka & 289 y & grain size; magnetic susceptibility  \\
Linxia loess \cite{yang2014}& 35.62, 103.2 & 2013 m  & 130 - 0 ka & & grain size \\
CHILOMOS loess stack \cite{yang2014}  & N/A& N/A & 249 - 0 ka & 200 y  & grain size
\\* \bottomrule
\caption{Loess and dust records, ordered by latitude.}
\label{tab:table4}\\
\end{longtable}

\begin{longtable}{@{}lllllll@{}}
\toprule
\textbf{Site name} & \textbf{Location} & \textbf{Elevation} & \textbf{Age} & \textbf{Res.} & \textbf{Proxies}  \\* \midrule
\endhead
Lake El'gygytgyn \cite{melles2012,meyer2014} & 67.5, 172.104 & 456 m & 3.6 - 0 Ma & 27 y  & Mn/Fe; Si/Ti; magnetic susceptibility; \\ & & & & & TOC; TIC; biogenic Si  \\
Lake Baikal \cite{prokopenko2006} & 53.696, 108.352  & 456 m & 1.8 - 0 Ma & 223 y  & biogenic silica \\
Füramoos \cite{muller2003} & 447.98333, 9.88333& 662 m & 140 - 0 ka  & & pollen  \\
Les Echets \cite{ampel2008,veres2009} & 45.8333, 5 & 267 m  & 46 - 15 ka  && pollen; diatoms; magnetic susceptibility; \\ & & & & & geochemistry  \\
Lac du Bouchet \cite{reille1990,sanchez2017}& 44.83, 3.82 & 1200 m  & 70 - 0 ka & 89 y & pollen; temperate forest pollen  \\
Summer Lake \cite{benson2003} & 42.83, -120.75& 1260 m  & 46 - 23 ka & 76 y & $\delta^{18}$O; $\delta^{13}$C  \\
Valle di Castiglione \cite{follieri1989,sanchez2017} & 41.9, 12.76 & 44 m & 56 - 14 ka & 357 y  & pollen; temperate forest pollen \\
Tenaghi Philippon \cite{tzedakis2006} & 41.17, 24.33& 40 m & 1.35 - 0 Ma  & & pollen \\
Lake Ohrid \cite{wagner2019,sadori2016,donders2021}& 41.049, 20.715  & 693 m & 1.36 - 0 Ma  & 208 y  & $\delta^{18}$O; $\delta^{13}$C; Zr/K; TIC; pollen; ...  \\
Lago di Monticchio \cite{allen1999,allen2000,huntley1999}  & 40.932, 15.605  & 656 m & 100 - 10 ka  & 120 y  & pollen; temperate forest pollen; \\ & & & & &  biogenic silica; dry density \\
Ioannina \cite{tzedakis2004} & 39.75, 20.85& 470 m & 130 - 0 ka & & pollen \\
Lake Van \cite{pickarski2017} & 38.667, 42.669 & 1649 m  & 250 - 129 ka & 336 y  & $\delta^{18}$O; $\delta^{13}$C; pollen; ...\\
Padul \cite{camuera2018} & 37, -3.67 & 785 m & 197 - 0 ka & & pollen \\
Dead Sea \cite{miebach2019} & 31.508, 35.471  & N/A & 88 - 14 ka & 242 y  & pollen \\
Lake Tulane \cite{grimm2006} & 27.584, -81.502  & 36 m & 61 - 0 ka & & Pinus pollen \\
Lake Tanganyika \cite{tierney2008} & -6.714, 29.833 & 773 m & 59 - 1 ka& 260 y  & $\delta^{13}$C; $\delta$D; Lake Surface Temperature \\
Lake Malawi \cite{johnson2016} & -11.294, 34.437& 500 m & 1.28 - 0 Ma  & 14 y & Ca; T; $\delta^{13}$C \\
Lake Titicaca \cite{fritz2007,fritz2010} & -15.937, -69.16  & 3810 m  & 370 - 3 ka & 28 y & TOC; $\delta^{13}$C; grain size
\\* \bottomrule
\caption{Lake sediment cores, ordered by latitude.}
\label{tab:table5}\\
\end{longtable}

\newpage

\section*{\centering\LARGE{Supplementary Figures}}
\renewcommand{\figurename}{Supplementary Figure S}

\section*{CENOGRID stack, including transitions detected using a shorter window length}

\begin{figure*}[!htbp]
\includegraphics[width=0.75\linewidth]{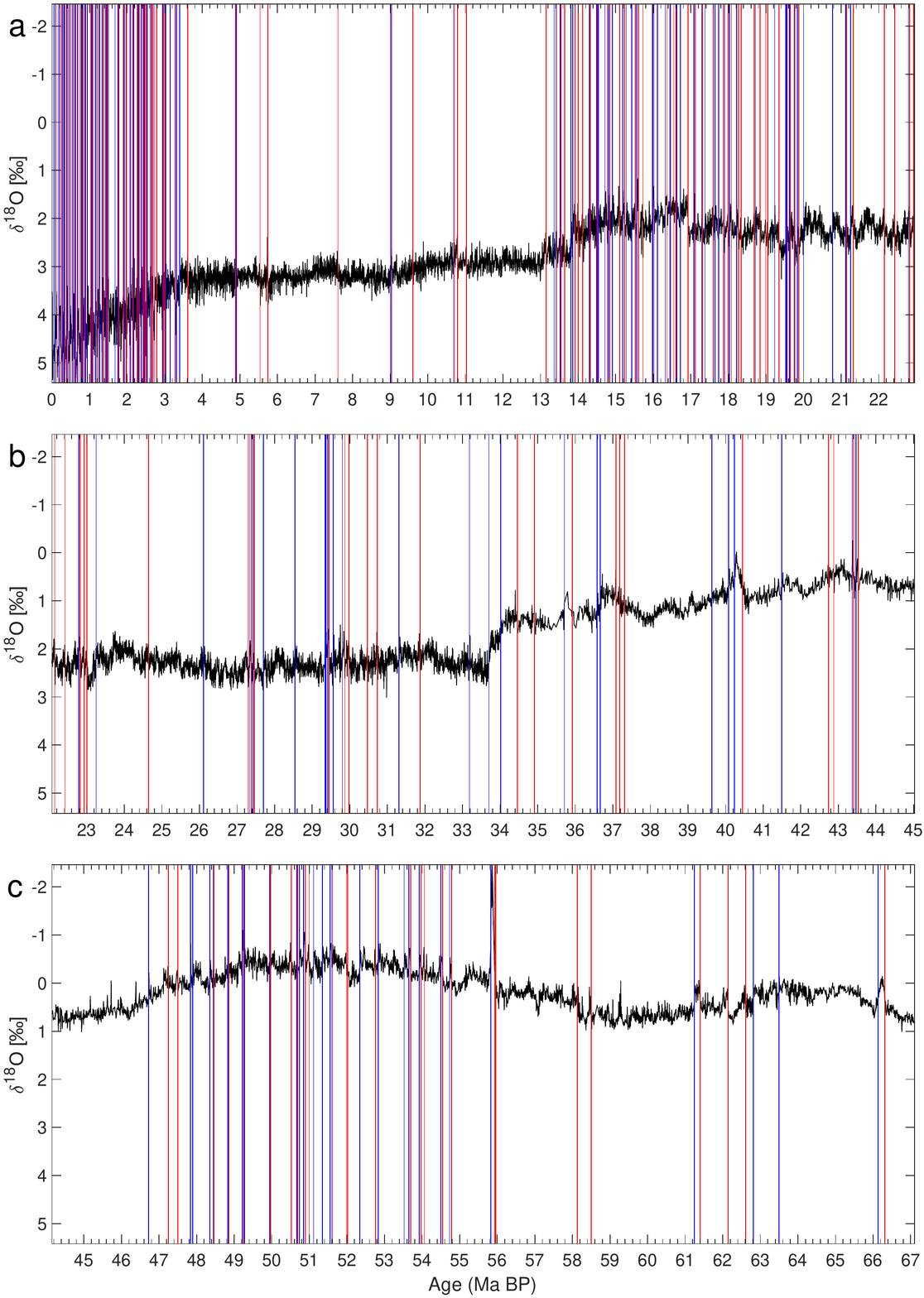}
    \centering\caption{CENOGRID stack of benthic $\delta^{18}$O \cite{westerhold2020}: (a) 0–22.9 Ma BP; (b) 22.1–45 Ma BP; and (c) 44.2–67.1 Ma BP. Vertical lines represent transitions detected by the KS test \cite{bagniewski2021}, with red lines for warming transitions and blue lines for cooling ones.
	Transitions detected for the entire record using a window length of $0.02 \le w \le 2.5$~Myr.
 The vertical axes are reversed.}
\label{cenogr}
\label{cenogr2}
\end{figure*}

\newpage

\section*{U1308 marine sediment core, including detected transitions}

\begin{figure*}[!htbp]
\includegraphics[width=0.8\linewidth]{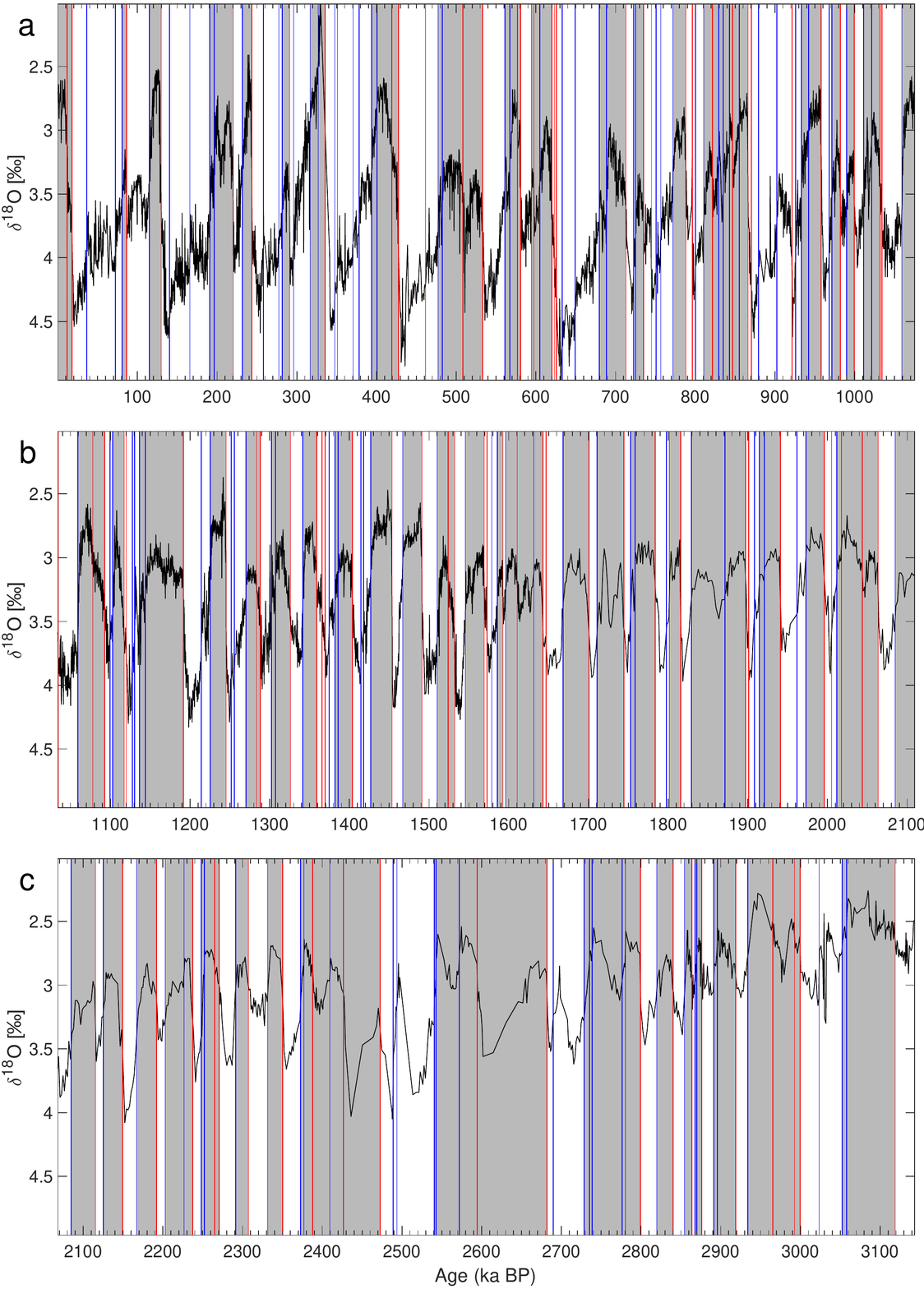}
    \centering\caption{Benthic \emph{Cibicidoides sp.} record in the U1308 marine sediment core $\delta^{18}$O \cite{hodell2016}: (a) 0–1.07 Ma BP; (b) 1.04–2.11 Ma BP; and (c) 2.07–3.14 Ma BP. Vertical lines represent transitions detected by the KS test \cite{bagniewski2021}, with red lines for warming transitions and blue lines for cooling ones. Vertical axes are reversed. Marine isotope stages (MISs) are shaded, with grey bars representing interglacials (odd-numbered MISs), while white bars represent glacials (even-numbered MISs).}
\label{U1308}
\end{figure*}

\newpage

\bibliography{supplement}